\documentclass[aps,prev,twocolumn,preprintnumbers,floatfix,nofootinbib]{revtex4-1}

\newcommand{\be}{\begin{equation}}
\newcommand{\ee}{\end{equation}}
\newcommand{\bea}{\begin{eqnarray}}
\newcommand{\eea}{\end{eqnarray}}

\usepackage[utf8]{inputenc}
\usepackage{graphics}
\usepackage{mathrsfs}
\usepackage{amsmath, amsthm, amssymb}
\usepackage[dvipsnames]{xcolor}
\usepackage{epstopdf}
\usepackage[pdftex]{graphicx}
\usepackage{amssymb}
\usepackage{verbatim}
\usepackage{hyperref}
\usepackage{enumerate}
\usepackage{float}
\usepackage[caption = false]{subfig}
\usepackage[normalem]{ulem}  

\begin{document}

\title{Axion Miniclusters Made Easy}

\author{David Ellis}
\email{david.ellis@uni-goettingen.de}
\author{David J. E. Marsh}
\email{david.marsh@uni-goettingen.de}
\author{Christoph Behrens}
\email{christoph.behrens@uni-goettingen.de}

\vspace{1cm}
\affiliation{Institut f\"{u}r Astrophysik, Georg-August Universit\"{a}t, Friedrich-Hund-Platz 1, D-37077 G\"{o}ttingen, Germany}

\begin{abstract}

We use a modified version of the Peak Patch excursion set formalism to compute the mass and size distribution of QCD axion miniclusters from a fully non-Gaussian initial density field obtained from numerical simulations of axion string decay. We find strong agreement with N-Body simulations at significantly lower computational cost. We employ a spherical collapse model, and provide fitting functions for the modified barrier in the radiation era. The halo mass function at $z=629$ has a power-law distribution $M^{-0.6}$ for masses within the range $10^{-15}\lesssim M\lesssim 10^{-10}M_{\odot}$, with all masses scaling as $(m_a/50\mu\mathrm{eV})^{-0.5}$. We construct merger trees to estimate the collapse redshift and concentration mass relation, $C(M)$, which is well described using analytical results from the initial power spectrum and linear growth. Using the calibrated analytic results to extrapolate to $z=0$, our method predicts a mean concentration $C\sim \mathcal{O}(\text{few})\times10^4$. The low computational cost of our method makes future investigation of the statistics of rare, dense miniclusters easy to achieve.

\end{abstract}

\maketitle

\section{Introduction}
    
    The quantum chromodynamics (QCD) axion \cite{PhysRevLett.38.1440,PhysRevLett.40.223,PhysRevLett.40.279,PhysRevLett.43.103,SHIFMAN1980493,SHIFMAN1980493,DINE1981199,Zhitnitsky:1980tq} is a hypothetical particle, the pseudo-Goldstone boson of a spontaneously broken global $U(1)$ symmetry, known as the Peccei-Quinn (PQ) symmetry. It is added to the Standard Model (SM) to solve the charge-parity (CP) problem of QCD \cite{PhysRevD.92.092003, Dar:2000tn, PhysRevD.98.030001}. The axion acquires mass from QCD instantons, with the zero temperature value $m_a=5.691(51)\mu \mathrm{eV} \frac{10^{12}\mathrm{GeV}}{f_a}$ \cite{Borsanyi:2016ksw,diCortona:2015ldu,Gorghetto:2018ocs}, where $f_a$ is the scale of PQ symmetry breaking divided by the colour anomaly. Axion couplings to the SM are suppressed by powers of $f_a$, leading to very small couplings and a long axion lifetime, making the axion an excellent dark matter (DM) candidate. If $f_a$ is restricted to be below the Planck scale, the allowed range of masses is $10^{-11}\text{ eV}\lesssim m_a\lesssim 10^{-2}\text{ eV}$, where the lower bound arises due to black hole superradiance~\cite{Arvanitaki:2014wva,Stott:2018opm} and the upper bound from the SN1987A neutrino burst~\cite{Raffelt:2006cw,Chang:2018rso}.
    
    The power spectrum of the axion in the early Universe, and its subsequent relic density, is critically dependent on when PQ symmetry breaking occurs in cosmic history. If the symmetry is broken before or during inflation, and is not restored later, our Universe begins with a uniform initial value for the axion field $\phi=f_a\theta$, with $\theta$ chosen randomly from a uniform distribution [$-\pi,\pi$]. In this case, cold axions are produced in the early universe through the non-thermal vacuum realignment mechanism~\cite{PRESKILL1983127,ABBOTT1983133,DINE1983137}, and the relic density depends on the random value of $\theta$. Depending on the level of fine tuning, essentially the entire axion mass range is consistent with the DM relic density, with intermediate values preferred if $\theta^2\sim 1$.
    
    In the alternative case, the PQ symmetry is broken during the normal thermal evolution of the Universe, and the model predictions are no longer sensitive to the UV physics of inflation. The production via the Kibble mechanism~\cite{Kibble_1976, KIBBLE1980183}, and subsequent decay by vacuum realignment, of axion strings and domain walls allows a direct computation of the relic abundance depending only on $m_a$. Due to the hierarchy of scales between $m_a$ and $f_a$, however, the numerical problem cannot be solved directly and there is no definite consensus on the value of $m_a$ required to give the correct relic abundance, even in the case with only strings and no long lived domain walls. It is reasonable to assume that $m_a\gtrsim 20 \,\mu\text{eV}$ is required in order not to over produce DM, with larger values preferred in the case with long lived domain walls~\cite{Armengaud_2019,Gorghetto:2018myk,Vaquero:2018tib,Buschmann:2019icd,Klaer:2017ond,Hiramatsu:2012sc,PhysRevD.85.105020,PhysRevD.83.123531,PhysRevLett.73.2954,HARARI1987361, PhysRevLett.124.021301}. 
    
    The decay of the defects also predicts large amplitude, small scale density perturbations, which are the seeds for \emph{axion miniclusters} \cite{HOGAN1988228, Kolb:1995bu, Kolb:1994fi}. If a significant fraction of DM is bound up in miniclusters, this could have a significant impact on direct detection experiments targeting the high mass regime suggested by the relic density from defect decay~\cite{PhysRevLett.118.091801,Marsh:2018dlj,Lee:2019mfy,Baryakhtar:2018doz,Mitridate:2020kly}. Impacts between axion miniclusters and the Earth have been estimated to be as rare as once every $10^5$ years, while tidal stripping of miniclusters in the Milky Way leads to a diffuse background and the possibility to detect tidal streams~\cite{Tinyakov:2015cgg,Dokuchaev:2017psd, OHare:2017yze, Lawson:2019brd}. On the other hand, miniclusters offer new opportunities for indirect searches such as gravitational microlensing \cite{Fairbairn:2017dmf,Fairbairn:2017sil, Kolb:1995bu}, and radio astronomy \cite{Foster:2020pgt,Tkachev:2014dpa}.
    
    It is thus crucial to be able to accurately predict the mass function, $dn/dM$, and size distribution of miniclusters. We propose to solve this problem accurately and efficiently using the Peak Patch formalism~\cite{Stein:2018lrh} applied to axion density fields computed from field theory solutions of topological defect decay~\cite{Vaquero:2018tib}. The Peak Patch formalism, as we will describe, is an application of the extended Press-Schechter model~\cite{1974ApJ...187..425P,1986ApJ...304...15B} for non-linear gravitational collapse, which solves the so-called excursion set on a real space realisation of the density field, thus accounting for all non-Gaussianity. Our approach is able to accurately reproduce the results of recent high resolution N-body simulations~\cite{Eggemeier:2019khm} at a fraction of the numerical cost, and extends and clarifies the successes of purely analytical models~\cite{Fairbairn:2017sil,Enander_2017}. It thus opens the door to the much further study of miniclusters.
    
The characteristic mass of axion miniclusters can be estimated by calculating the total mass in axions contained within the horizon of the universe when the axion field first began to oscillate at $T_{1}\approx1$ GeV. This point is given by 
\begin{equation}
    AH(T_{1}) \approx m_a(T_{1}).
\end{equation}
where $A$ is, somewhat arbitrarily, normally chosen to be 1 or 3. The temperature dependence of the axion mass arises due to the susceptibility of the topological charge $\chi(T)$ via the equation
\begin{equation}
    m_a^2 = \frac{\chi(T)}{f_a^2}.
\end{equation}

A power law dependence on temperature $\chi(T) \propto T^{-b}$ is predicted by dilute instanton gas approximation with $b = 8.16$. However, this approximation can be avoided by determining $\chi(T)$ directly using lattice QCD \cite{Borsanyi:2016ksw}.

The radius of the horizon at $T_1$ is given by
\begin{equation}
    L_{1} = \frac{1}{a_{1}H(T_1)}
\end{equation}
From this, the characteristic minicluster mass can be estimated using 
\begin{equation}
    M_{\textsc{mc}} = \bar{\rho}_a(T_{1})V_{\textsc{h}}L_1^3,
    \label{Mmc}
\end{equation}
where the value $V_{\textsc{h}}$ varies between different definitions of the Hubble volume, with the definition of Ref.~\cite{Kolb:1995bu} corresponding to $V_{\textsc{h}}=1$, while e.g. Refs.~\cite{Fairbairn:2017sil,Dai:2019lud} use a larger value $V_{\textsc{h}}\approx 130$. The result is well fit by
\begin{equation}
    M_{\textsc{mc}} = 7.36 \times 10^{-12}  \Big( \frac{T_{1}}{\mathrm{ GeV}} \Big)^{-3} \mathcal{S}\left(\log_{10}\frac{T_1}{\text{ GeV}}\right) V_{\textsc{h}} M_{\odot},
\end{equation}
\begin{align}
\begin{split}
    \mathcal{S}(x) \equiv 0.5[1 + \tanh 4(x+0.8)] \\ +   1.3[1 + \tanh 4(-0.8 - x)],
\end{split}
\end{align}
where $\mathcal{S}(x)$ is an activation function which encodes the effect of the QCD phase transition impact on $g_*(T)$, and varies from $\mathcal{S}\approx 1$ for $T_1\gtrsim 1\text{ GeV}$ to $\mathcal{S}\approx 2.6$ for $T_1\lesssim 1\text{ GeV}$.

We consider a reference axion mass of $m_a = 50\mathrm{\mu eV}$ which is related to a characteristic mass of $M_{\textsc{mc}} \approx 4.5\times10^{-13} M_{\odot}$. However, as will be seen, the true minicluster mass function is in fact very wide. Therefore, characteristic mass should only be considered a very approximate value.

Going beyond this simple estimate requires many ingredients, and can be done via many different methods \cite{Enander:2017ogx, Fairbairn:2017sil}. Those without gravity have all to date used arbitrary overdensity thresholding \cite{Vaquero:2018tib, Buschmann:2019icd, Kolb:1995bu}. Additionally, Press-Schechter (PS) methods have only looked at very late redshifts and have not considered realistic initial conditions from simulation, including the effects of the strong non-Gaussianities caused by the decay of topological defects.  

Typically, dark matter simulations employ N-Body methods to calculate the formation and evolution of structure. However, such simulations are computationally expensive and it is therefore unlikely that they could ever be used to calculate the full evolution of axion MCs from their formation deep in the radiation dominated era to today. Additionally, these methods fail to offer a theoretical understanding for these processes of structure formation. Hence there is potentially much to be gained from developing more analytical approach to understanding minicluster formation. In this paper we discuss the application of both analytical and semi-analytical approaches, in the forms of a Press-Schechter analysis and the mass Peak Patch formalism respectively, to better understand these problems.

This paper is organised as follows; First, in Sec. ~\ref{sec:MiniSeeds} we briefly outline the PQ mechanism and describe the field theory simulations performed in Ref.~\cite{Vaquero:2018tib} which are used to calculate the initial density field used in this work. Then, in Sec. ~\ref{sec:GravCollapse} we outline three different gravitational collapse models; the non-linear collapse of a spherical matter overdensity in Sec. ~\ref{sec:Spherical}, the linear growth of density perturbation in Sec. ~\ref{sec:LinearCollapse} and the Press-Schechter formalism in Sec. ~\ref{sec:PS}. Next, in Sec. ~\ref{sec:ConcEst}, we outline a formalism for estimating the concentration parameter of a dark matter halo from its collapse redshift. This is then applied analytically from the power spectrum, under the assumption of Gaussianity, to predict the concentrations of axion miniclusers from our initial conditions, fitting to the N-Body results from Ref.~\cite{Eggemeier:2019jsu} at $z=99$. We subsequently relax the assumption of Gaussianity by using Peak Patch formalism in Sec. ~\ref{sec:PeakPatch}. Using this formalism we calculate the minicluster mass function as well as building merger trees from which we can estimate the collapse redshift and hence concentration of each of our miniclusters at $z=99$. Finally, we extend our estimate using the power spectrum to $z=0$ to predict the concentrations of axion miniclusters today.

\section{Minicluster Seeds}
\label{sec:MiniSeeds}

The PQ mechanism is achieved by introducing a complex scalar given by
\begin{equation}
    \psi = \chi e^{i \phi/f_a},
\end{equation}
where $\phi$ is the axion field. This field has a U(1) shift symmetry which is broken by the scalar potential
\begin{equation}
    V(\psi) = \lambda \Big( |\psi|^2 - \frac{f_a^2}{2} \Big)^2.
\end{equation}
Due to this potential, as the universe cools the complex field takes on it's vacuum expectation value $\langle \psi \rangle = f_a/\sqrt{2}$.  This spontaneous symmetry breaking causes the initial phase of the field (i.e. the axion) to be a random distribution in causally disconnected regions. This consequently produces a network of cosmic strings and domain walls \cite{KIBBLE1980183, Kibble_1976}. 

The axion field then continues to evolve as a Klein-Gordon equation on in an expanding Friedman-Robertson-Walker metric as given by
\begin{equation}
    \Ddot{\theta} + 3H\theta - \frac{1}{a^2}\nabla^2\theta + \frac{1}{f_A^2}\partial_{\theta}V_{\mathrm{QCD}}(T,\theta) = 0.
\label{eq:AxField}
\end{equation} 
The potential is given by $V_{\mathrm{QCD}}\approx \chi(T)(1 - \cos\theta)$. This produces an additional phase transition leads to a globally preferred value of the axion field causing it to oscillate at $AH(T)\approx m_a(T)$ as discussed previously the leading to the decay of the topological defects \cite{Vaquero:2018tib}. 

\begin{figure}
\includegraphics[width=0.99\columnwidth, trim=0 20 0 0, clip]{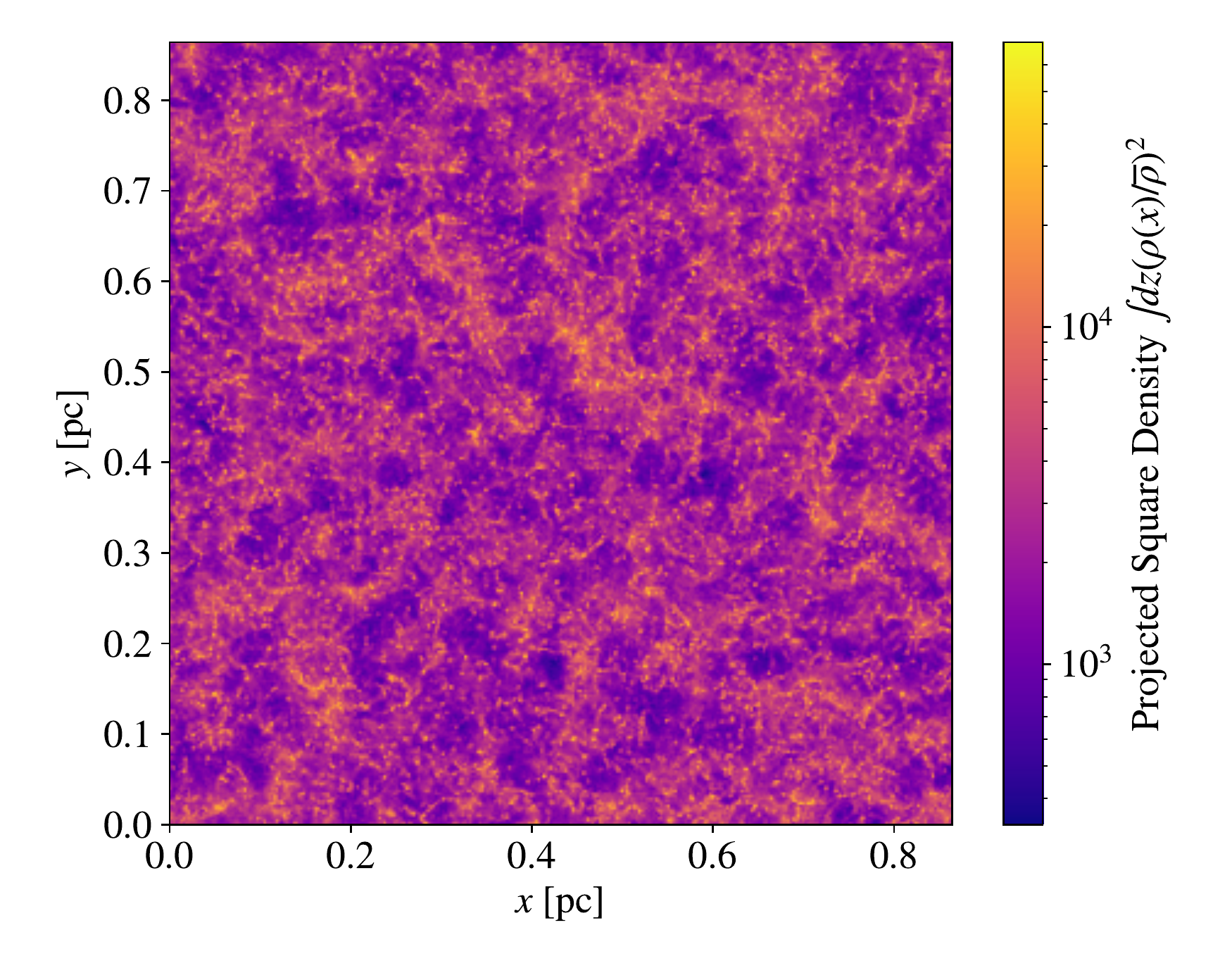}
\caption{\emph{Projected Square Density of Axion DM Initial Conditions.} Field at $z=10^6$, calculated using field theory simulations from Ref. \cite{Vaquero:2018tib}. The scale of the projected densities is set by the average square density along the line of sight.} 
\label{fig:InitCond}
\end{figure}

\begin{figure}[ht!]
\includegraphics[width=0.99\columnwidth, trim=0 10 0 0, clip]{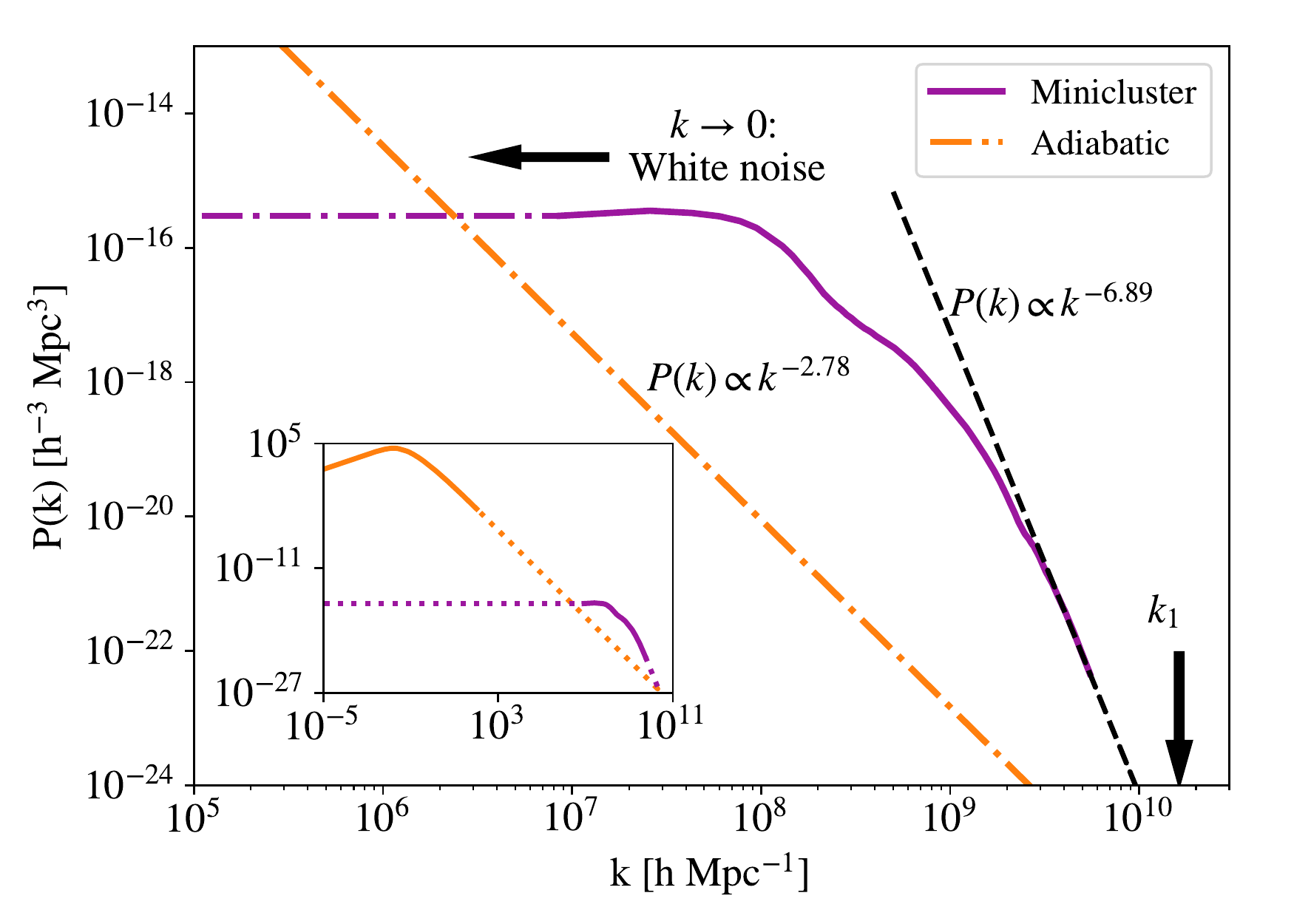}
\caption{\emph{Power Spectrum of Initial Conditions.} Data from Ref. \cite{Vaquero:2018tib}. The extrapolated adiabatic power spectrum (orange dot-dashed) is included for comparison. The adiabatic power will also be cut-off below $k_{1}$, since this is the axion Jeans scale. The inset graph shows the full extent of these two power spectra as well as the points at which their extrapolated curves meet. The adiabatic power spectrum is computed with CAMB and extrapolated with the given power law to large $k$ \cite{Lewis:1999bs}.}
\label{fig:PowerSpec}
\end{figure}

Our initial conditions are taken from recent field theory simulations of Ref.~\cite{Vaquero:2018tib} in which the PQ field is evolved through the QCD phase transition, using the Press-Ryden-Spergel method \cite{1989ApJ...347..590P, PhysRevD.65.023503}. These simulations include the formation of axion strings and $N=1$ domain walls. The defects decay when the axion field begins to oscillate, at which point the radial field is at the vacuum expectation value everywhere and so can be switched off. The resulting distribution of the axion field at $z\approx10^6$ is shown in Fig.~\ref{fig:InitCond}. We show the projected square density, which allows us to see the large density perturbations which seed axion miniclusters throughout the entire length of the box. The power spectrum $P(k)$, where  $k$ is Fourier conjugate to $x$, of this field, linearly evolved to $z=0$, is shown in Fig.~\ref{fig:PowerSpec}.

The simulations are performed in code units which absorb the axion mass (allowing us to freely consider any value). The characteristic time scale is defined as using $H(T_1) = m_a(T_1)$ ($A=1$). The comoving coherence length of the axion field is used to define a characteristic length scale given by
\begin{equation}
    L_1 = 0.0362 \Big(\frac{50\mathrm{\mu eV}}{m_a} \Big)^{0.167} \mathrm{pc}.
\end{equation}
where the exponent of 0.167, valid in the 1 to 2 GeV temperature range of interest, is found by fitting using $\chi(T)$ \cite{diCortona:2015ldu, Borsanyi:2016ksw}. The total box size is then given by $L_{\mathrm{box}}=L_{\mathrm{adm}}L_1$ where $L_{\mathrm{adm}}$ is an integer which dictates the specific size of the box. In our case $L_{\mathrm{adm}} = 24$ giving a comoving box length of 0.864 pc.\footnote{This box is somewhat larger than those studied originally in Ref.~\cite{Vaquero:2018tib}, and were provided to us privately. They are the same as used in the N-Body simulations of Ref. \cite{Eggemeier:2019khm}.} Since this is the only occurrence of the axion mass, as will be seen later, choosing a different value of $m_a$ simply shifts the final minicluster mass range.

\section{Gravitational Collapse}
\label{sec:GravCollapse}

\subsection{Spherical non-linear collapse}
\label{sec:Spherical}
As done in Ref.~\cite{Kolb:1994fi}, we consider a spherical matter overdensity of radius $r$ in a universe containing only matter and radiation. The equation of motion for such a region can be written as
\begin{equation}
    \ddot{r} = -\frac{8\pi G}{3}\rho_r - \frac{G M_{\textsc{tot}}}{r^2},
\end{equation}
where $\rho_r$ is the homogeneous radiation density and $M_{\textsc{tot}}$ is the mass enclosed by the radius. We consider a comoving reference frame $r = a(\eta)R_{\xi}(\eta){\xi}$ where $R$ is the deviation from the Hubble flow and $\xi$ is a label for the comoving shell. Then, in a flat $\Omega_0 = 1$ universe containing only radiation and pressureless matter it can be shown the equation of motion can be rewritten as 
\begin{equation}
    x(1+x)\frac{\mathrm{d}^2R}{\mathrm{d}x^2} + \Big(1 + \frac{3}{2}x\Big) +\frac{1}{2}\Big(\frac{1+\delta_{\mathrm{i}}}{R^2} - R\Big) = 0,
    \label{eq:KTeom}
\end{equation}
where $x = a/a_{eq}$ is the normalised scale factor and $\delta_{\mathrm{i}} = \frac{\rho - \bar{\rho}}{\bar{\rho}}$ is the clump overdensity \cite{Kolb:1994fi}. This can then be solved numerically, assuming $R_i = 1$ and $\frac{dR}{dx}\Big|_i = 0$. The density of the matter perturbation at turn around is given by
\begin{equation}
    \langle\rho_{\textsc{ta}}\rangle = \frac{1}{4\pi r^2}\frac{\mathrm{d}M}{\mathrm{d}r}.
\end{equation}
We therefore represent the parameters at turn around using
\begin{equation}
    x_{\textsc{ta}} = \frac{C_x}{\delta_{\mathrm{i}}},
\end{equation}
and
\begin{equation}
    \langle\rho_{\textsc{ta}}\rangle = C_{\rho}\rho_{eq}\frac{\delta_{\mathrm{i}}^3}{3\xi^2}\frac{d}{d\xi}(1+\delta_{\mathrm{i}})\xi^3,
\end{equation}
where $C_{\rho} = 1/(R_{\textsc{ta}}C_x)^3$ depends only weakly on $\delta_i$. From the virial theorem, we can approximate that the virial radius is half the turn around radius. Therefore, for the core density $\delta_{\mathrm{i,0}} = \delta_{\mathrm{i}}(r=0)$ we have
\begin{equation}
    \langle\rho_{\textsc{f}}\rangle= 8C_{\rho}\rho_{\mathrm{eq}}\delta_{\mathrm{i,0}}^3(1+\delta_{\mathrm{i,0}}).
    \label{eq:FinalDens}
\end{equation}
Therefore, by numerically solving the collapse of $R$ we can calculate $C_{\rho}$ and hence estimate the final density. We will see that for large overdensities (large $\delta_{\mathrm{i}}$) that the final overdensity is strongly sensitive to the redshift at which we consider the object to first start collapsing. 

We have confirmed, as found in Ref.~\cite{Kolb:1994fi}, that $C_{\rho}$ depends only very weakly on $\delta_{\mathrm{i,0}}$ with a value of $C_{\rho} \sim 17$ thus giving 

\begin{equation}
    \langle\rho_{\textsc{f}}\rangle= 136\rho_{\mathrm{eq}}\delta_{\mathrm{i,0}}^3(1+\delta_{\mathrm{i,0}}).
\end{equation}

\subsection{Linear growth} \label{sec:LinearCollapse}

Considering small perturbations in an expanding homogeneous and isotropic universe the equation of motion describing the linear growth of matter perturbations is given by 
\begin{equation}
    \frac{\mathrm{d}^2\delta}{\mathrm{d}t^2} + 2H\frac{\mathrm{d}\delta}{\mathrm{d}t}-\frac{c_s^2}{a^2}\Delta\delta - 4\pi G \epsilon_0 \delta = 0,
\end{equation}
where $c_s$ is the speed of sound. For cold matter in the presence of radiation or dark energy, we can neglect the $c_s^2$ term since the pressure is negligible. It can then be shown that the equation of motion can be written as 
\begin{equation}
    x^2(1+x^{-3w})\frac{\mathrm{d}^2\delta}{\mathrm{d}x^2} + \frac{3}{2}x(1+(1-w)x^{-3w})\frac{\mathrm{d}\delta}{\mathrm{d}x}- \frac{3}{2}\delta = 0,
\end{equation}
where $w$ is the equation of state for the homogeneous relativistic background energy with $x$ defined as before \cite{mukhanov_2005}. Hence, for dark matter in the presence of a radiation background
\begin{equation}
    x(x+1)\frac{\mathrm{d}^2\delta}{\mathrm{d}x^2} + (1+\frac{3}{2}x)\frac{\mathrm{d}\delta}{\mathrm{d}x}-\frac{3}{2}\delta = 0.
    \label{Meszaros}
\end{equation}
This is known as the Meszaros equation, and is equivalent to Eq.~\ref{eq:KTeom} considering $R\equiv1-\delta$ in the limit $\delta \ll 1$. For the case of interest, $w=1/3$, the general solution is
\begin{equation}
\begin{split}
    \delta(x) &=  C_1\Big(1+\frac{3}{2}x\Big) \\ 
    &+ C_2\Big[\Big(1 + \frac{3}{2}x\Big)\ln\frac{\sqrt{1+x}+1}{\sqrt{1+x}-1} - 3\sqrt{1+x}\Big],
\end{split}
\end{equation}
where $C_1$ and $C_2$ are constants which are fixed by the initial conditions. 

Since we are considering a single isocurvature mode in the sub-horizon limit the initial conditions are determined by
\begin{align}
    \delta(t=0) &= 1, \\
    \frac{d\delta}{dt}\Big|_{t=0} &= 0.
\end{align}
However, since
\begin{equation}
    \frac{d\delta}{dt} = \frac{d\delta}{dx}H(x)x = \frac{d\delta}{dx}\dot{x},
\end{equation}
and $H(x)$ and $x$ are both nonzero, so we require
\begin{equation}
    \frac{d\delta}{dx} = 0.
\end{equation}
In the limit of small $x$ 
\begin{equation}
    \delta(x) = (C_1 - 3C_2) - C_2\ln\Big(\frac{x}{4}\Big) + \mathcal{O}(x),
\end{equation}
therefore $C_1 = \delta_i$, $C_2 = 0$ \cite{mukhanov_2005}. This allows us to write the growth factor simply as 
\begin{equation}
    D(x) = 1 + \frac{3}{2}x.
    \label{eq:GrowthFactor}
\end{equation}
By calculating the linear overdensity at the time of collapse, as calculated using the non-linear equation of motion, we can calculate the size of regions that should collapse as a function of redshift. 

Solving Eq ~\ref{eq:KTeom} numerically we find, similarly to the turnaround parameter of Ref.~\cite{Kolb:1994fi}, the collapse parameter $C_{\mathrm{col}}(\delta_i)\equiv\delta_i x_{\mathrm{col}}(\delta_i)$ is approximately constant as a function of the initial overdensity. We can therefore expect an overdensity to collapse once it has grown by some fixed amount $\delta_{\mathrm{c}}(\delta_i) = \delta_i+\varepsilon$.

To ensure that the $\delta_i$ dependence of $C_{\mathrm{col}}$ is correctly dealt with, we fit $\varepsilon$ to our numerical solution using a least squares fit and find the familiar value of $\varepsilon=1.686$.

\begin{figure}
\includegraphics[width=\columnwidth, trim=0 20 0 0, clip]{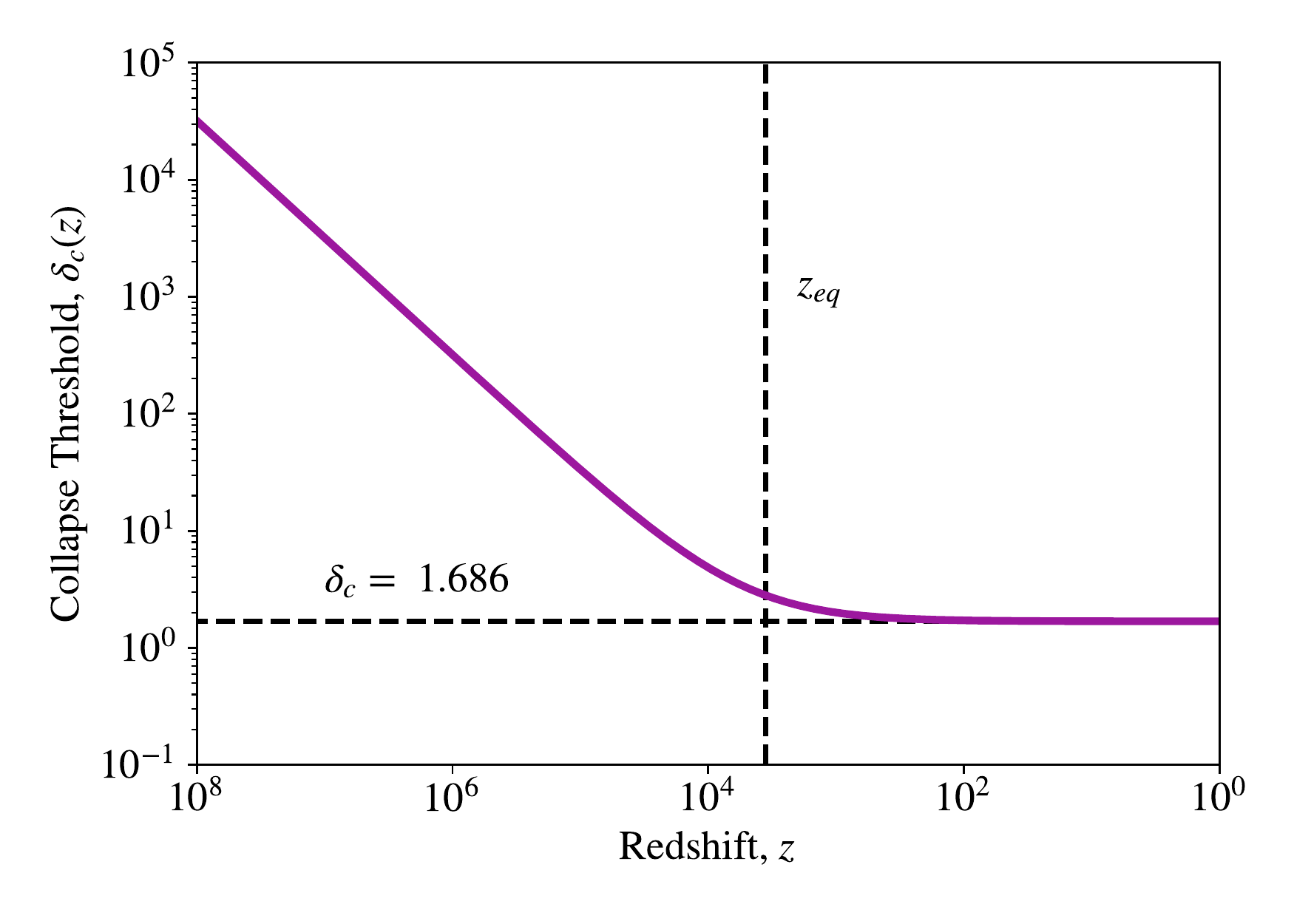}
\caption{\emph{Modified critical overdensity.} Calculated by solving the spherical collapse as given by Eq. ~\ref{eq:KTeom} to determine the point of non-linear collapse and relating this to density from linear growth using Eq. ~\ref{eq:GrowthFactor}. Performing a least-squares fit, we find that The collapse threshold saturates to the Einstein-de Sitter value 1.686 at late times, and is larger during the radiation era.} 
\label{fig:DensityThresh}
\end{figure}

Then, since $\delta_i = \delta_c(x)/D(x)$ the critical density can be calculated as a function of redshift. Hence, the critical overdensity above which structures will collapse at some redshift $z$ is then given by
\begin{equation}
    \delta_{\mathrm{c}}(x) = \frac{\varepsilon D(x)}{D(x)-1}.
    \label{eq:Thresh}
\end{equation}
It can be seen, as shown in Fig.~\ref{fig:DensityThresh}, that this modified threshold reproduces the usual CDM result of $\delta_c=1.686$ in the limit of small redshifts. Using this equation we can relate the linear growth of overdensities to non-linear collapse.

\subsection{Press-Schechter}
\label{sec:PS}

From the spherical collapse model we expect that regions of a linearly evolved overdensity field above the critical overdensity will have collapsed to form virialised objects. This can be written as the criteria that $\delta(\textbf{x}, t)>\delta_c(z)$. To assign masses to these regions, we consider a smoothed density field given by
\begin{equation}
    \delta_s(\textbf{x};R) \equiv \int \delta(\textbf{x}')W(\textbf{x}+\textbf{x}';R)\mathrm{d}^3\textbf{x}',
\end{equation}
where $W(\textbf{x};R)$ is a window function of characteristic radius R. For the simplest case of a spherical tophat filter, this is related to a mass $M=\frac{4}{3}\pi\bar{\rho}R^3$. The PS formalism works by assuming that the probability of $\delta_s>\delta_c(t)$ is equal to the fraction of mass contained within halos of mass greater than $M$ \cite{1974ApJ...187..425P}. Assuming a Gaussian density field (we address non-Gaussianities shortly), this probability is found to be
\begin{equation}
    \mathscr{P}[>\delta_c(t)]=\frac{1}{2} \mathrm{erfc} \Big[\frac{\delta_c(t)}{\sqrt{2}\sigma(M)}\Big],
\end{equation}
in which $\sigma(M)$ is the mass variance of the smoothed density field given by 
\begin{equation}
    \sigma^2(M)=\langle \delta_s^2(\textbf{x};R)\rangle = \frac{1}{2\pi^2}\int^{\infty}_{0}P(k)\Tilde{W}^2(\textbf{k}R)k^2\mathrm{d}k,
    \label{eq:MassVarience}
\end{equation}
where $P(k)$ is the power spectrum of the density perturbations and $\Tilde{W}(\textbf{k}R)$ is the window function in Fourier space. For a tophat filter, the window function is given by
\begin{equation}
    \Tilde{W}_{\mathrm{th}}(\textbf{k}R)=\frac{3(\sin{kR}-kR\cos{kR})}{(kR)^3}.
    \label{eq:TopHatWindow}
\end{equation}
However, underdense regions can be enclosed within a larger overdense region causing them to be included in the larger collapsed object. This cloud-in-cloud problem is accounted for by introducing a famous 'fudge factor' of 2 \cite{1991ApJ...379..440B}. This modifies the original ansatz to give the following
\begin{equation}
    F(>M)=2\mathscr{P}[>\delta_c(t)].
\end{equation}
From this, this halo mass function (HMF) can be estimated by
\begin{equation}
    \frac{\mathrm{d}n}{\mathrm{d}\ln M} = M \frac{\rho_0}{M^2}f(\sigma) \Big| \frac{\mathrm{d}\sigma}{\mathrm{d}\ln M}\Big|,
    \label{eq:HMF_PS}
\end{equation}
where $\rho_0$ is the mean DM density and $f(\sigma)$ is the multiplicity function.

In standard PS, this multiplicity function is given by
\begin{equation}
    f_{\textsc{ps}}(\sigma) = \sqrt{\frac{2}{\pi}}\frac{\delta_c}{\sigma}\exp\Big(\frac{-\delta_c^2}{2\sigma^2} \Big)
    \label{eq:f_PS}
\end{equation}
where $\delta_c$ is the critical overdensity. As discussed, for ordinary WIMP CDM $\delta_c \approx 1.686$, however we replace this with out redshift dependent threshold $\delta_c=\delta_c(z)$. There exist, however, other multiplicity functions which are often found to better match simulation data than PS \cite{Murray:2013qza}. One of the most popular alternatives is the Sheth-Tormen fit
\begin{equation}
    f_{\textsc{st}}(\sigma) = A\sqrt{\frac{2a}{\pi}}  \Big[1 + \Big( \frac{\sigma^2}{a\delta_c^2} \Big)^p \Big]\frac{\delta_c}{\sigma}\exp\Big(\frac{-a\delta_c^2}{2\sigma^2} \Big),
    \label{eq:f_ST}
\end{equation}
where $A = 0.3222$, $a = 0.707$ and $p = 0.3$ \cite{10.1046/j.1365-8711.2001.04006.x}.

Applying the outlined formalism on the power spectrum of the initial conditions, using one of the fitting functions above, we can estimate the HMF as a function of redshift assuming a Gaussian overdensity field.

\subsection{Estimating the Concentration Parameter}\label{sec:ConcEst}

It is known from previous simulations of the same initial density field that the halos formed have a radial density profile that is well described by the usual NFW profile
\begin{equation}
    \rho(r) = \frac{\rho_c}{(r/r_s)(1+r/r_s)^2},
    \label{eq:NFW}
\end{equation}
where $\rho_c$ and $r_c$ are the characteristic density and radius respectively \cite{Eggemeier:2019jsu, Eggemeier:2019khm, Navarro:1995iw}. The concentration parameter is then defined by $c = r_{200}/r_s$ where $r_{200}$ is the radius which encloses a region of mean density equal to $\rho_{200} = 200\rho_{\mathrm{crit}}$.

Efforts have been made using N-Body simulations to relate the concentration parameter of a halo to it's collapse redshift \cite{10.1093/mnras/275.3.720, Navarro:1995iw, Navarro:1996gj, Bullock:1999he}. 

Navarro et. al defined the collapse redshift, sometimes also called the formation redshift, to be the redshift at which for some halo half of its final mass $M_{\mathrm{final}}$ is contained within progenitors of a mass larger than $fM_{\mathrm{final}}$ where $f$ is some fraction. This can be estimated using the Press-Schechter formalism
\begin{equation}
    \mathrm{erfc} \big\{ X(z_{\mathrm{col}}) - X(z_{\mathrm{0}}) \big\} = \frac{1}{2},
    \label{eq:PS-zcol}
\end{equation}
where
\begin{equation}
    X(z) = \frac{\delta_c(z)}{\sqrt{2[\sigma^2(fM, z) - \sigma^2(M, z)]}},
\end{equation}
and $\sigma(M, z)$ is the linear variance of the power spectrum at some final redshift $z$ \cite{10.1093/mnras/262.3.627}. Navarro et.  al found that a fraction of $f=0.01$ best matched the results from N-Body simulations of CDM with adiabatic initial conditions. We take f as a free normalisation constant in our application.

Solving Eq.~\ref{eq:PS-zcol} as a function of $z_0$ for a fixed mass by using the mass variance for the powerspectrum shown in Fig. \ref{fig:PowerSpec}, we find that the collapse redshift initially falls proportional to $z_0$, then at $z_0 \sim z_{eq}$ the collapse redshift tends towards some constant value determined by the final halo mass $M$.

\begin{figure}
\includegraphics[width=\columnwidth, trim=0 20 0 0, clip]{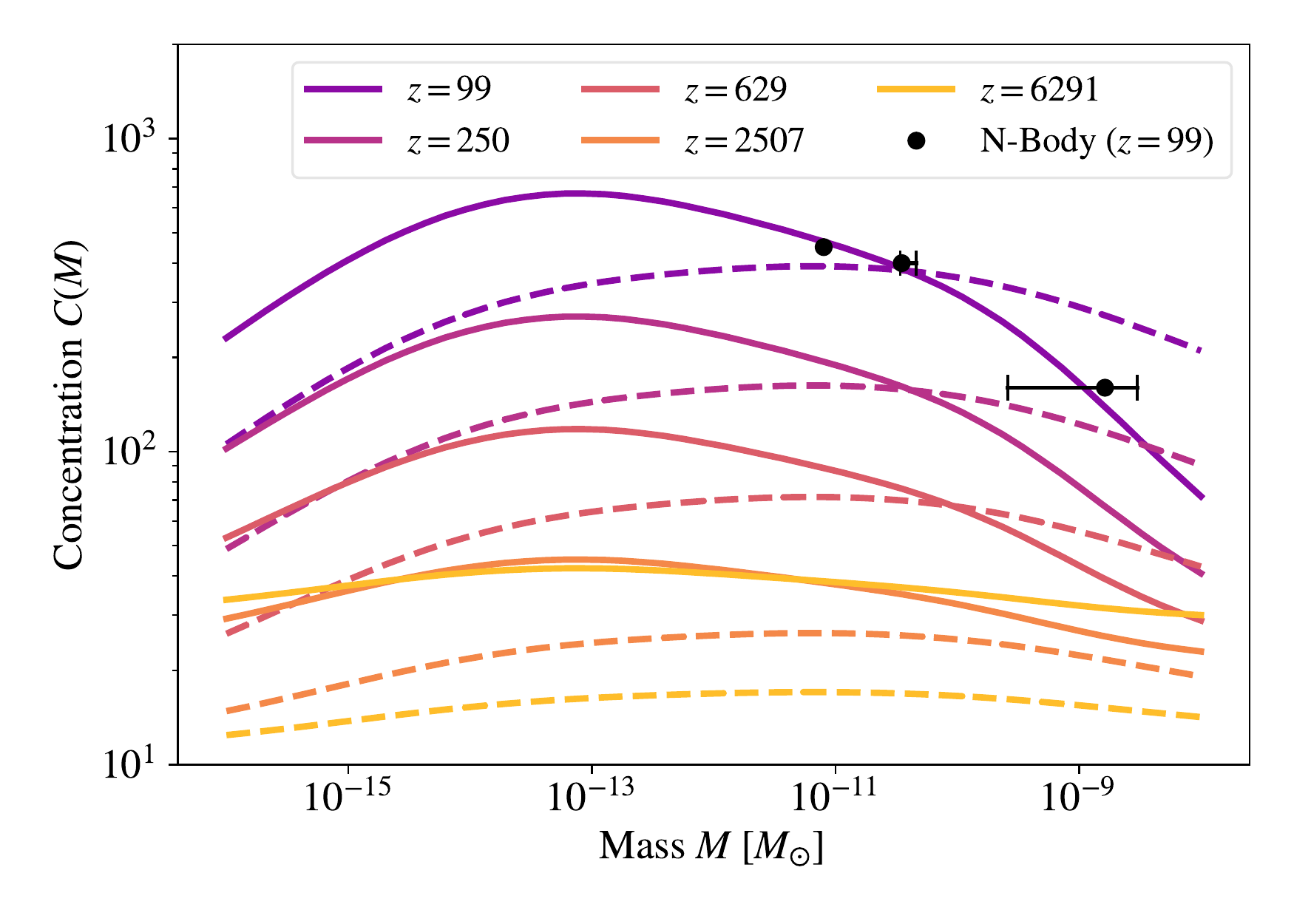}
\caption{\emph{Concentration  Parameter  Estimates}. Curves calculated by solving Eq. ~\ref{eq:PS-zcol} using the initial power spectrum to estimate the collapse redshift and combining with the concentration estimation formalism of NFW (solid lines) via Eq. \ref{eq:NFWConc1} and Eq. \ref{eq:NFWConc2} using $f=0.01$  and the concentration estimation formalism of Bullock et. al (dashed lines) via Eq. \ref{eq:BullockConc} with $f=10^{-5}$. The free parameters are determined using a least-squares fit to N-Body results from Ref. \cite{Eggemeier:2019khm} and are found to be $\kappa_{\textsc{nfw}} = 1.28 \times 10^5$ and $\kappa_{\textsc{b}} = 10.7$ respectively.} 
\label{fig:PS_conc}
\end{figure}

They then assume that the scale density of the halo is proportional to the ratio of the density of the universe at this time to the density today. Thus implying
\begin{equation}
    \delta_s(x, \kappa) = \kappa_{\textsc{nfw}}(f) \frac{\rho_u(x_\mathrm{col})}{\rho_u(x)},
    \label{eq:NFWConc1}
\end{equation}
where $\kappa_{\textsc{nfw}}$ is a free parameter and $x_\mathrm{col}$ is the normalized scale factor relating to the point of collapse defined by Navarro et. al (not to be confused with the point of non-linear collapse from Sec ~\ref{sec:LinearCollapse}). Since the halos they considered only collapse at very late redshifts only the matter component of the density was considered. However, since the halos we consider collapse much earlier, we must also acknowledge the contribution of radiation. Therefore, we take
\begin{equation}
    \rho_u(x_\mathrm{col}) = \rho_{\mathrm{eq}}( x_\mathrm{col}^{-3} + {x_\mathrm{col}}^{-4}).
    \label{eq:rho_u}
\end{equation}
Having calculated the characteristic overdensity, the concentration parameter can then be calculated by solving
\begin{equation}
    \delta_s = \frac{200}{3}\frac{c^3}{\ln(1+c) - c/(1+c)}.
    \label{eq:NFWConc2}
\end{equation}
Once the concentration parameter curve is calculated, the free parameter $\kappa_{\textsc{nfw}}(f)$ can be fitted to results from N-Body simulations. In Fig.~\ref{fig:PS_conc} we fit the predicted concentration parameter from the NFW formula with the power spectrum in Fig.~\ref{fig:PowerSpec} to the N-Body results from \cite{Eggemeier:2019khm} and find $\kappa_{\textsc{nfw}}(f=0.01) = 1.28 \times 10^5$. We can then calculate the predicted concentration parameter as a function of redshift, as shown as solid lines in Fig. ~\ref{fig:PS_conc}. The error bars seen on the N-Body values denote the range of halo masses used to calculate the concentration parameter. This was done by averaging the spherical density profiles and then fitting the resulting profile to Eq.~\ref{eq:NFW}.

We see that at very early redshifts, all halos are predicted to have approximately the same concentration. This is because the threshold for collapse is still very high, as seen from Fig.~\ref{fig:DensityThresh}. As a result, any collapsed object that is found, regardless of mass, will have most likely only just formed. Then, since the concentration is assumed to depend only on the collapse redshift, the value is therefore the same for all masses.

At around matter-radiation equality a peak begins to form at $M \approx 10^{-13}M_{\odot}$. The location of this peak does not change over time. This indicates that at any point in time, objects with a mass around this value are the oldest and therefore most concentrated. 

More recently, Bullock et. al developed a more general model \cite{Bullock:1999he}. Firstly they simplify their definition of the collapse redshift to be the redshift at which for some halo half of its final mass $M_{\mathrm{final}}$ is contained within progenitors of any mass. This is equivalent to setting $f=0$ in the NFW definition.

Additionally, they generalise the model by defining a more general characteristic density $\tilde{\rho_s}$, defined by $M_{\mathrm{vir}} \equiv \frac{4 \pi}{3}r_s^3\tilde{\rho_s}$. As done previously, this is then assumed to be associated with the density of the universe at the time of collapse according to
\begin{equation}
    \tilde{\rho_s} = \kappa_{\textsc{b}}^3\Delta_{\mathrm{vir}}(a)\rho_u(a),
\end{equation}
where $\kappa_{\textsc{b}}$ is a proportionality constant which represents the contraction of the inner halo in excess of the standard requirements from dissipationless top-hat halo virialistation. 

For a universe dominated only by matter it can then be found that
\begin{equation}
    c(M,a) = \kappa_{\textsc{b}}\frac{a}{a_{\mathrm{col}}} = \kappa_{\textsc{b}}\frac{x}{x_{\mathrm{col}}}.
    \label{eq:BullockConcUnmod}
\end{equation}
However, again this must be modified to account for early radiation domination through Eq. ~\ref{eq:rho_u}. The new relation is found to be 
\begin{equation}
    c(M,x) = \kappa_{\textsc{b}}\Big(\frac{x^{-3}+x^{-4}}{x_{\mathrm{col}}^{-3}+x_{\mathrm{col}}^{-4}}\Big)^{1/3}.
    \label{eq:BullockConc}
\end{equation}
It can be seen that we recover Eq.~\ref{eq:BullockConcUnmod} in the limit of large $x$ and $x_{\mathrm{col}}$. Since $x_{\mathrm{col}}$ tends towards a constant value, we can therefore predict, as done by Bullock et. al, that at late times $c(M, x) \propto x$. While it isn't quite as easy to see, the same is also true for the NFW case. 

As with the NFW approach, we use this to predict the concentration parameter. For consistency with the Bullock approach we use $f=10^{-5}\approx0$. Then, using a least-squares fit to the N-Body data at $z=99$ and find a proportionality constant $\kappa_{\textsc{b}} = 10.7$. This can be seen in as dashed curves in Fig.~\ref{fig:PS_conc}.

We see that the peak in concentration forms at $M \approx 2.4\times 10^{-11} M_{\odot}$, roughly two orders of magnitude larger than predicted using the NFW approach. It should be noted that the overall shape of the curve is determined only by the choice of the progenitor fraction $f$ (as used in Eq. ~\ref{eq:PS-zcol}). The different treatments of the scale density change primarily the normalisation of the curve at each redshift. 

It is clear from Fig.~\ref{fig:PS_conc} that the NFW approach fits the N-Body results the best. We therefore choose to apply this approach moving forwards to our solution of the excursion set on the full initial density field.

\section{Peak Patch and the Excursion Set}
\label{sec:PeakPatch}

Typical full three-dimensional N-Body dark matter simulations can be very computationally expensive. This expense is increased when simulated axion miniclusters since it begins to collapse deep into the epoch of radiation domination, thus requiring significantly longer simulation times. The mass-Peak Patch algorithm is an alternative to N-Body simulations which is able to generally reproduce the same results using a fraction of the CPU time and memory. We use a modified version of this algorithm produced in \cite{Stein:2018lrh}. 

\begin{figure}
\includegraphics[width=\columnwidth, trim=0 20 0 21, clip]{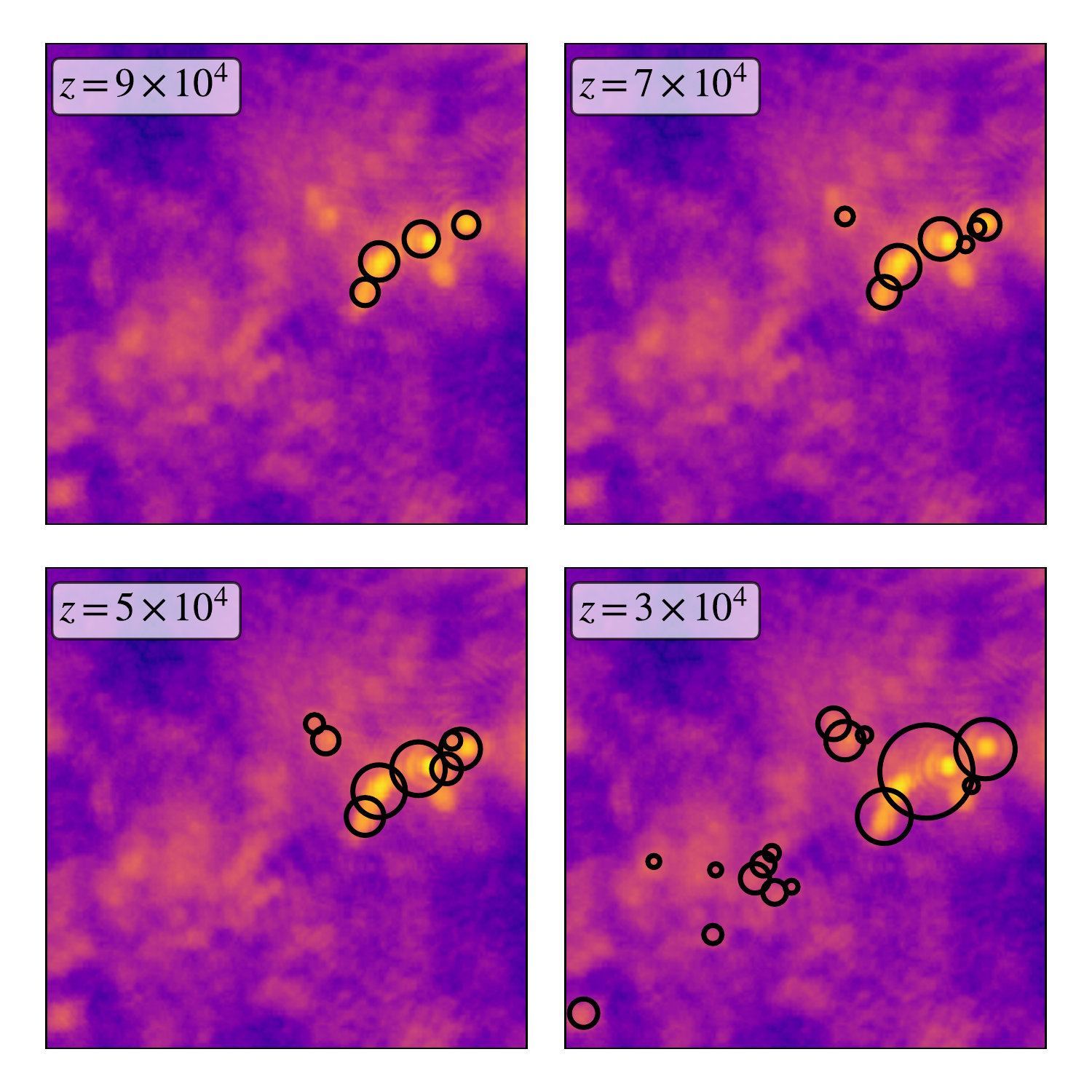}
\caption{\emph{Evolution of DM Halos calculated by Peak Patch.} Calculated for a small section overlayed onto the initial density field. Black circles indicate the radii of regions in which the contained overdensity is equal to the critical overdensity as shown in Fig.~\ref{fig:DensityThresh}. The section shown is a few percent of the total box size.}
\label{fig:PatchEvo}
\end{figure}

\begin{figure*}
\includegraphics[width=1.8\columnwidth, trim=0 22 0 0, clip]{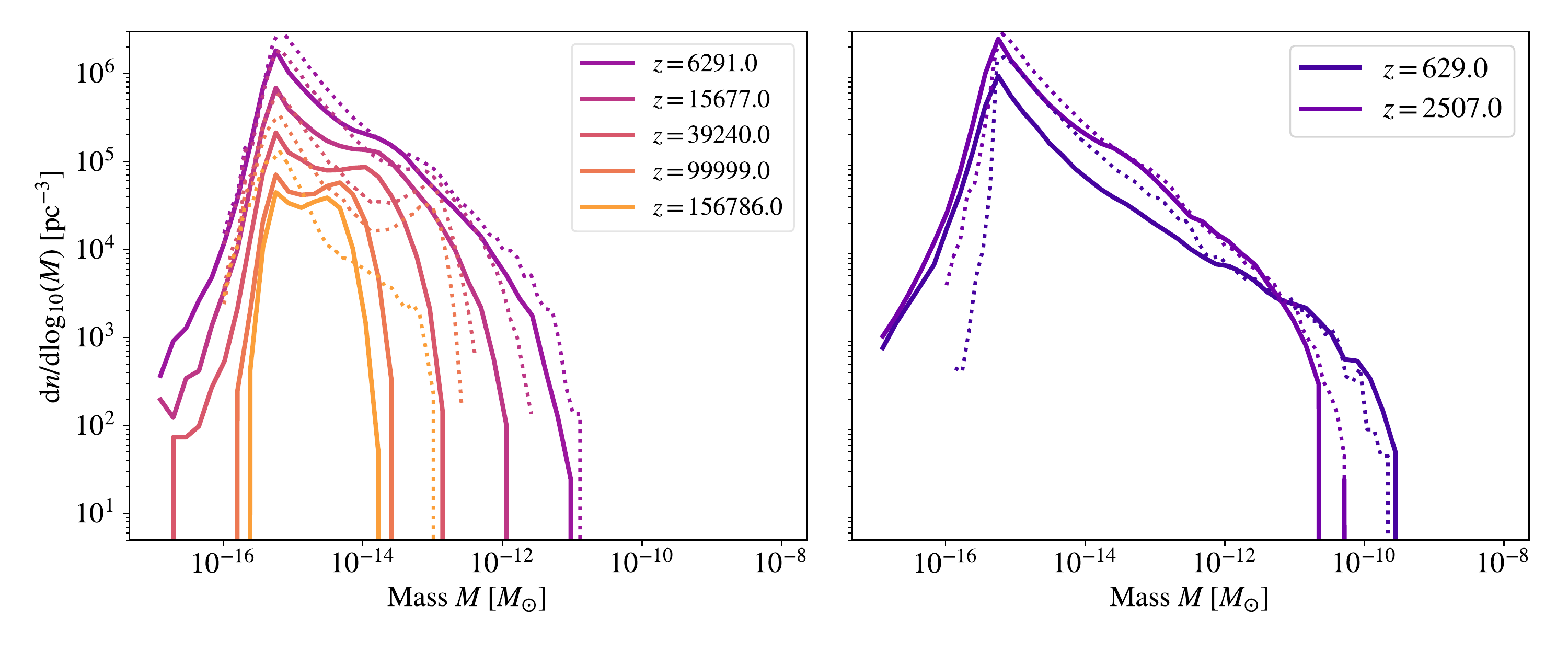}
\caption{\emph{Minicluster mass function from Peak Patch.} Solid lines show the HMF calculated using peak-patch with a redshift dependent collapse threshold calculated in \ref{sec:LinearCollapse}. Dotted lines show the HMFs at the same redshifts calculated using N-Body techniques in Ref.~\cite{Eggemeier:2019khm}.} 
\label{fig:AxionHMF}
\end{figure*}

Peak Patch is based on an extended Press-Schechter model in that it solves the excursion set in real space on a true realisation of the density field. Excursion sets are regions in which the density field exceeds the threshold for collapse. However, while a region of unfiltered space might be bellow this threshold, the same region can be found to be over-dense when filtered (smoothed) on some scale. Therefore, we have to filter the density field on a hierarchy of scales $R_f$ producing a four-dimensional field $F(\textbf{r}, R_f)$. When this filtering is performed in Fourier space with a sharp k-filter, the excursion set can be solved analytically reproducing the Press-Schechter HMF without requiring the same fudge factor. However, when filtering in physical space, this can not be calculated analytically as it leads to a correlated random walk. Therefore, numerical methods such as this one must be employed \cite{1991ApJ...379..440B}.

The Peak Patch  program utilises a massively-parallel procedure to identify peaks in a linearly evolved initial density field which has been smoothed on a range of scales. 
Once these peaks have been identified, a spherical integration is performed on the density field filtered on the largest scale at which the peak was found until the overdensity enclosed by the sphere is equal to the threshold overdensity at the redshift in question. This gives both the mass and the turn-around radius of the halos.
The final sizes of the halos are calculated using the spherical collapse approximation. The final set of halos is then calculated by excluding overlapping patches. From this the mass function can be calculated at each redshift. Additionally, the final locations of each peak are then calculated using first order Lagrangian displacements. These displacements are estimated by performing an integral over density field in Fourier space \cite{Stein:2018lrh}.

In it's unmodified form, Peak Patch identifies potential peaks as filtered cells containing an overdensity greater than the typical value of $\delta_{\mathrm{crit}} = 1.686$. The advantage for the present application is that, unlike PS, this method accounts correctly for all the non-Gaussianities in the initial distribution. A demonstration of the Peak Patch procedure is shown in Fig.~\ref{fig:PatchEvo}. Here we plot the location and extent of halos calculated by Peak Patch at four different redshifts on top of the projected square density field for a small region of our box.

To use this formalism to simulate the collapse of axion miniclusters, two primary modifications have been made. Firstly, instead of using a random initial field we input the realistic initial conditions calculated using field simulations by \cite{Vaquero:2018tib}. Secondly, we implement our redshift dependent threshold for collapse and growth factor from Eq. ~\ref{eq:Thresh} and Eq. ~\ref{eq:GrowthFactor} respectively. Using this we can now calculate the final sizes, masses and positions of the DM clumps originating from the realistic initial conditions.

We use a total of 28 real space filters, logarithmically spaced from $1.77\times10^{-3}$ pc to 0.229 pc to calculate the masses, sizes and positions of halos at 100 redshifts between $z=10^6$ and $z=99$. This final redshift was calculated in Ref.~\cite{Eggemeier:2019khm} to be safely before the box goes non-linear for our initial conditions. However, analysing the evolution of the power spectrum presented in Ref.~\cite{Eggemeier:2019khm}, we have determined that the largest scales deviate from linear growth after $z = 629$. We therefore conclude that beyond this redshift their simulation enters the quasi-linear regime on the scale of the simulation box. The fact that the largest scales are no longer linear calls into question the accuracy of the $N$-body results for $z<629$. In the quasi-linear regime power is ``moving'' across scales, and on the largest scales this is affected by the unphysical periodic boundary conditions.

Our results from Peak Patch depart significantly from the $N$-body results at $z<629$. This is caused by the same scales experiencing different growth (Peak Patch only uses linear growth),\footnote{If we artificially adopt the $N$-body quasi-linear growth at low redshift, this improves the agreement.} and results in a suppression in the number of low mass halos as a small number of large halos come to dominate the total box size. We therefore conservatively only consider the HMF to a minimum redshift of $z=629$, where the box scale is strictly linear.

However, while the halo number statistics are significantly affected for $z<629$, the concentrations are affected less so. This is because, as discussed, the $C(M)$ normalisation follows the growth, and the linear and quasi-linear growth only depart by $\mathcal{O)}(10\%)$ even at $z=99$. This can be understood by considering that the formation redshifts of surviving low mass halos (those not swallowed up by spurious large halos) are not affected; $C(M)$ is not a number statistic. Therefore, we still fit the concentrations of the halos by Peak Patch to those simulated by  Ref.~\cite{Eggemeier:2019khm} at $z=99$, this also being the only redshift at which such a comparison can be made.

In Fig.~\ref{fig:AxionHMF} we present the calculated HMFs at seven different redshifts together with the results from Ref.~\cite{Eggemeier:2019khm} in which the same initial conditions are simulated using traditional N-Body techniques.

Qualitatively the results are very similar. Both methods show that the number of objects at all masses increase rapidly over time up to around matter-radiation equality where the distribution falls and broadens to higher masses. We understand this evolution as the initial collapse of high density regions followed later by the merging of existing halos. At ($z=629$), we find that the HMF has a slope of $M^{-0.6}$, this is slightly lower than the value of $M^{-0.7}$ calculated by N-Body. This difference could be a hint at the box beginning to leave the linear-regime as it is consistent with the fact that our method finds more high-mass halos at this redshift. These halos are potentially suppressing the number of low-mass halos as they take up a dominant fraction of the total matter within the box.

\begin{figure}
\includegraphics[width=\columnwidth, trim=0 20 0 0, clip]{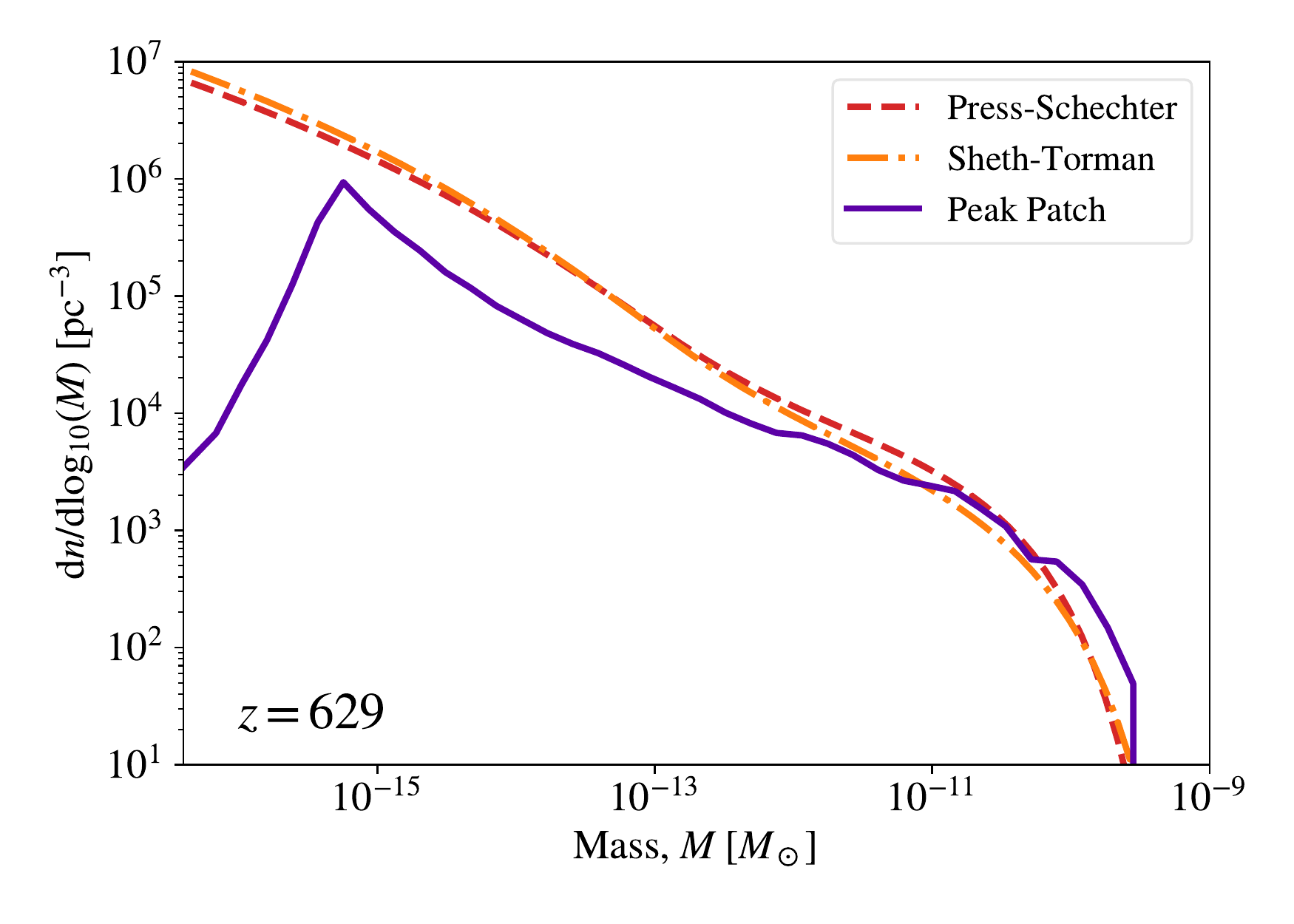}
\caption{\emph{Comparison of HMF with PS Prediction.} Solid line indicates the HMF calculated using peak-patch at $z=629$. Dotted and dot-dashed lines show the Press-Schechter predictions for the HMF at the same redshift using fitting functions PS and ST.} 
\label{fig:PScomp}
\end{figure}

We compare the different analytical HMFs calculated assuming Gaussianity and the initial P(k) in Fig.~\ref{fig:PScomp}. We find that the standard Press-Schechter and Sheth-Torman fitting functions predict very similar HMFs. However, the Peak Patch calculation again produces fewer low-mass halos than predicted using this formalism. Since we can extrapolate the power spectrum to larger scales, the Press-Schechter estimate is immune to problems of going ``non-linear'' faced by Peak Patch and N-Body calculations. This therefore further suggests that the reduced number of small-halos in our Peak Patch estimate is due to the sizes of the largest halos becoming comparable with the total box volume.

\begin{figure}
\includegraphics[width=\columnwidth, trim=0 25 0 0, clip]{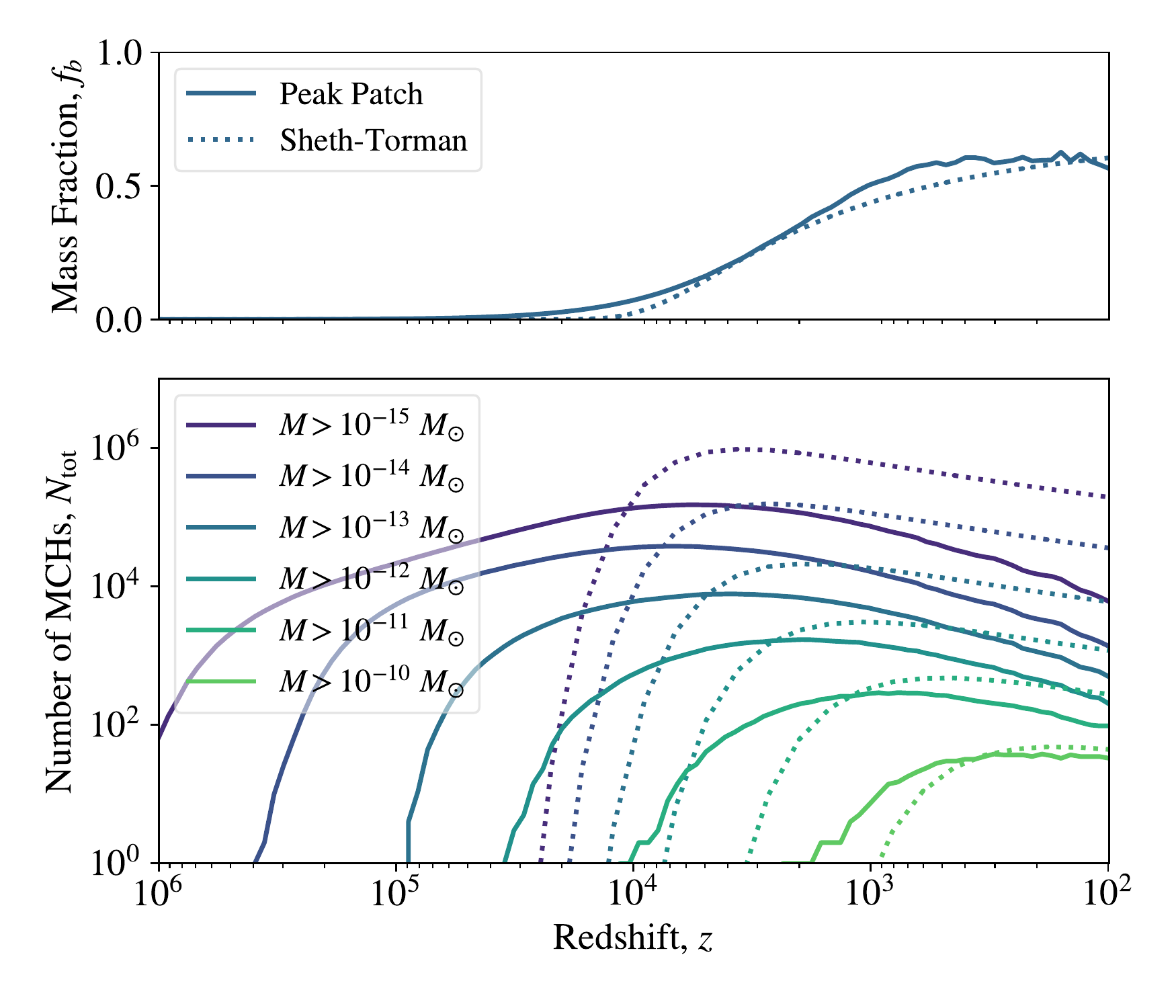}
\caption{\emph{Number of Minicluster Halos.} Top: Total fraction of  gravitationaly bound mass. Bottom: Total number of miniclusters above mass scales in the range $\{10^{-15}, 10^{-10}\} M_{\odot}$. Solid lines indicate the values calculated using peak-patch while dotted lines show the values predicted using Press-Schechter. Peak Patch finds a large number of halos to form at high redshift from the strongly non-Gaussian density peaks, consistent with the results of N-body.}
\label{fig:NumberofHalos}
\end{figure}

In Fig.~\ref{fig:NumberofHalos} we compare the calculated number of halos above different masses as a function of redshift to the value predicted from PS using the standard PS multiplicity function. For each mass range we see a very rapid initial growth of halos followed by a slow  stagnation and then decline. Interestingly, Peak Patch finds many more low mass objects at early redshifts than are predicted by PS. The N-Body simulations also find objects to form faster than PS predicts, however, these objects are still seen much later than in Peak Patch. This is though to be related to the fact that at early times our density field is highly non-Gaussian whereas the PS approach assumes exactly the opposite to be true.

\begin{figure}[h!]
\includegraphics[width=\columnwidth, trim=0 210 30 80, clip]{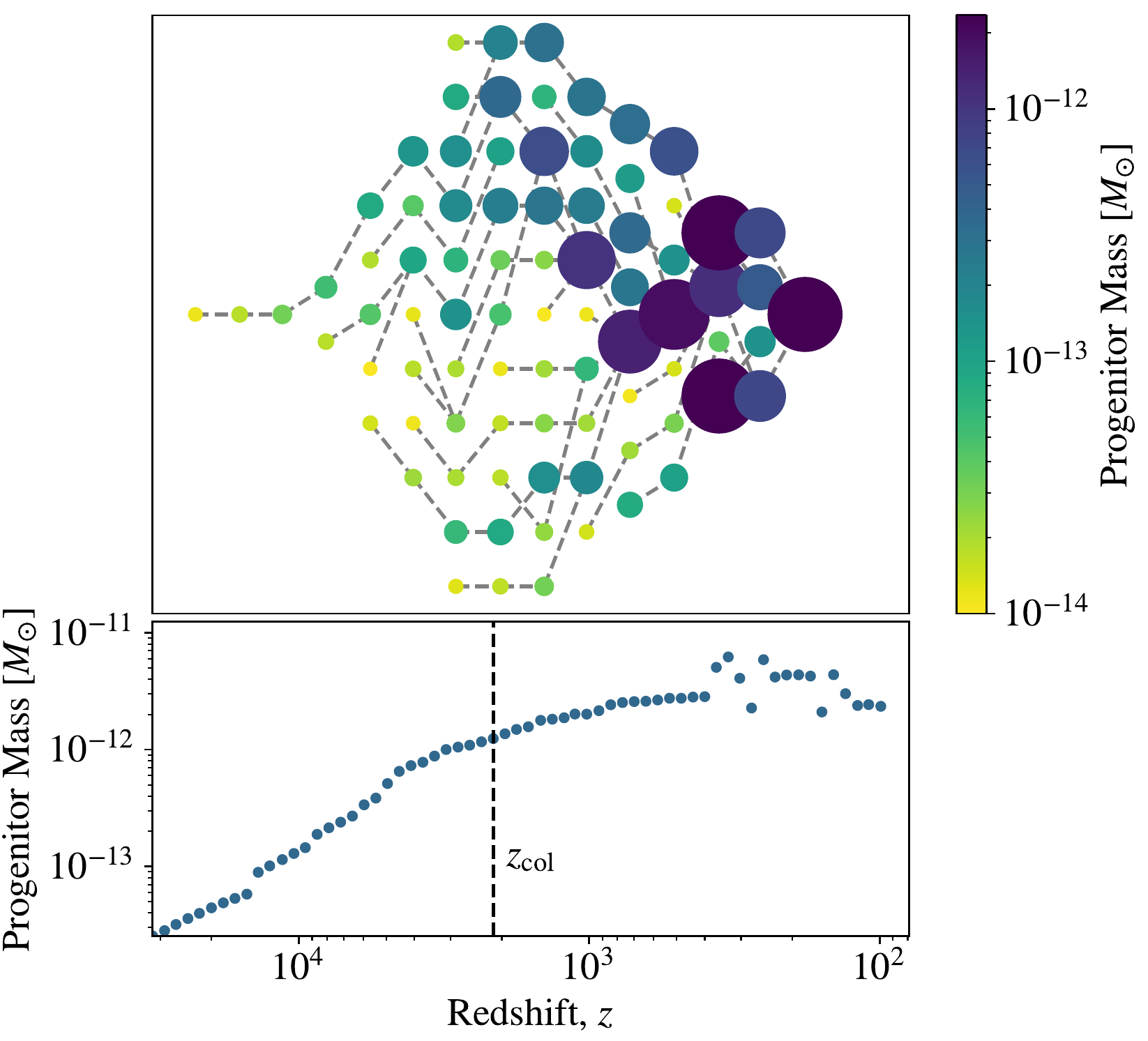}
\caption{\emph{Example merger tree}. Top: Merger tree for a halo with a final mass of $2.35 \times 10^{-12} M_{\odot}$ showing all progenitors with a mass greater than $10^{-14} M_{\odot}$. The radius of the circles is proportional to the effective radius of the progenitor. Bottom: Total mass of progenitors that have a mass greater that 0.01 times the final halo mass. The collapse redshift is then defined as the point at which this total equals half of the final mass, as introduced by Navarro et. al in Ref~.\cite{Navarro:1996gj}.} 
\label{fig:MergerTree}
\end{figure}

Next, we use the results from Peak Patch to build merger trees. An example case of a halo with a final mass of $\sim 10^{-12} M_{\odot}$ is shown in Fig.~\ref{fig:MergerTree}. 

The trees are built by choosing a halo from the data for the latest redshift, then, for each of the earlier redshifts, finding objects within the radius of the final halo plus the radius of the largest halo at that redshift. We then check for any regions overlapping with the final sphere. Following this, we calculate the volume of the overlapping region to assign an effective mass and radius to the sub-halo. Once a full list of peaks at each redshift is built, we can then assign sub-peaks to their parents by iteratively checking which peaks lie within the radius of the parent at the next latest redshift. 

It can be seen from example case shown that initially isolated halos grow through the accretion of surrounding matter. Then, there is a period in which many mergers take place. At this point the total mass of the progenitors stagnates and increases only slowly through accretion and occasional mergers.

Using these merger trees it is then possible to assign a collapse redshift to each halo using the half-mass definition as discussed in Sec.~\ref{sec:ConcEst}. In doing this we can then also estimate the shape of the concentration-mass curve from the true initial density field. The normalisation of the curve is then set by the same fitting parameter of $\kappa_{\textsc{nfw}} = 1.28 \times 10^5$ calculated earlier using Press-Schechter. 

The collapse redshift was calculated for the halos found by Peak Patch using three a threshold fraction of $f = 0.01$ however we found that, unlike with the PS predictions, the choice of threshold fraction only makes a significant impact for masses above the peak mass of $M\approx10^{-11}M_{\odot}$.

\begin{figure}[h!]
\includegraphics[width=\columnwidth, trim=0 20 0 15, clip]{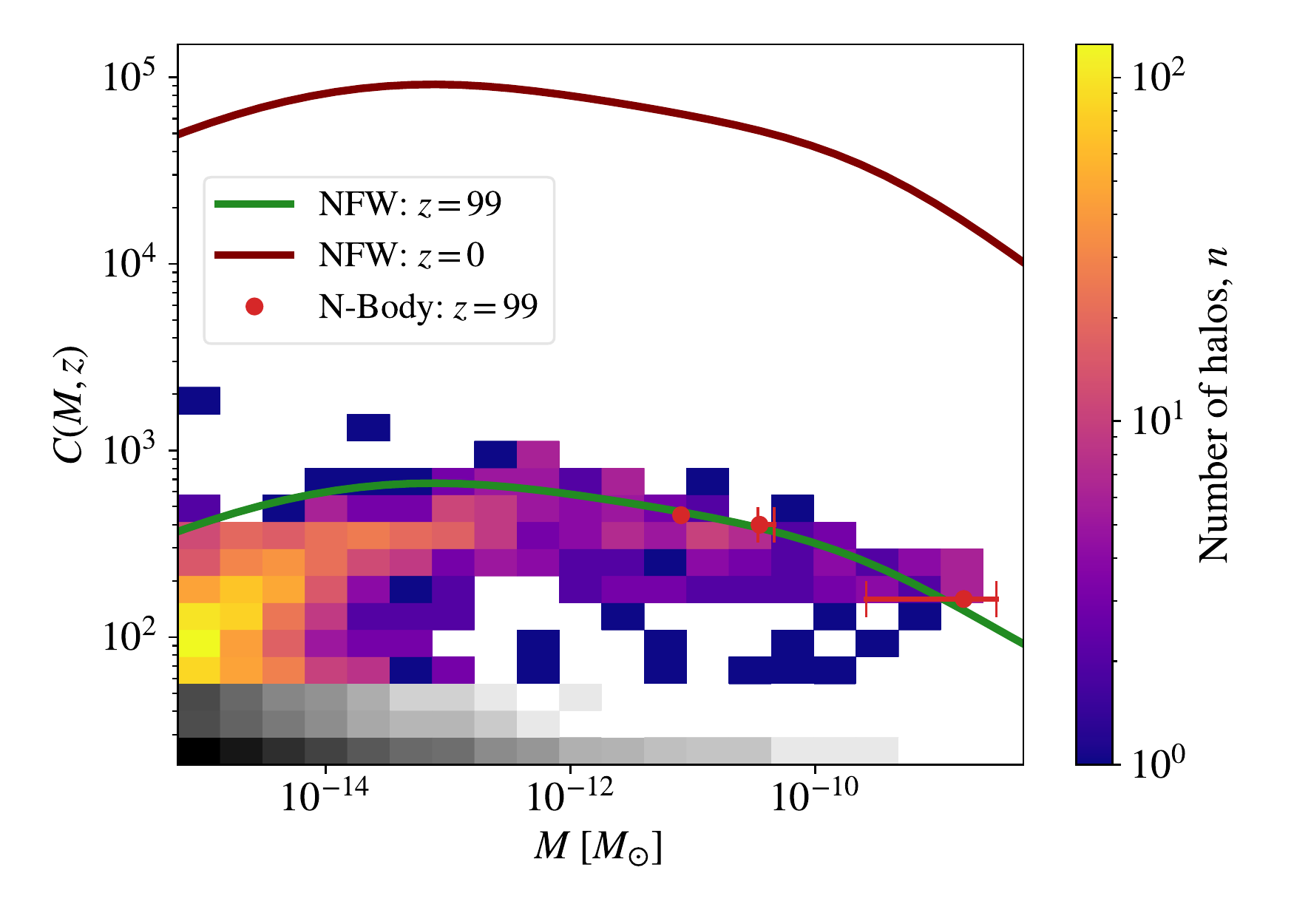}

\caption{\emph{Minicluster Concentrations}. Colour map indicates the concentration parameters calculated using Peak Patch merger trees. The solid green and dashed red lines show the concentrations from the power spectrum, assuming Gaussianity. The grey colour map shows halos calculated to have concentrations less than 50. Since these objects have only just formed it is thought that they have not had enough time to become as dense as the NFW prediction.}
\label{fig:finalConc}
\end{figure}

Using the same fitting parameter calculated for using the PS predictions for $C(M)$, we find a concentration distribution as shown in Fig.~\ref{fig:finalConc}. It can be seen that for masses above $\sim 10^{-12} M_{\odot}$ the Press-Schechter and Peak Patch approaches make comparable predictions for a large proportion of the miniclusters. However, Peak Patch also produces a large number of halos with a very low concentration at all masses up to $\sim 10^{-11} M_{\odot}$. 

Finally, given the success of our calibrated analytical approach in fitting both the N-body and Peak Patch results, we can extend our PS calculation to $z=0$ to predict the concentration parameter of the axion miniclusters today. Doing so, we find a maximum value of $7.9 \times 10^4$ at $M = 2.0 \times 10^{-13}$ as shown in Fig.~\ref{fig:finalConc}.

\section{Discussion and Conclusions}

We have presented analytic and semi-analytic methods to compute the mass function and halo concentration of axion miniclusters, using the simulated initial conditions from Ref.~\cite{Vaquero:2018tib}. 

Our main results use the Peak Patch excursion set formalism~\cite{Stein:2018lrh} to solve the barrier crossing and gravitational spherical collapse problem in real space for the fully non-Gaussian initial data. The resulting minicluster mass function displays initial growth during the radiation era due to direct collapse of rare overdensities and a falling critical barrier. In the matter dominated era, the barrier becomes constant in redshift, and hierarchical mergers take over. The qualitative behaviour is in exact agreement with results from N-body simulations~\cite{Eggemeier:2019khm}. 

At late times, the mass function is well described by the Sheth-Tormen~\cite{10.1046/j.1365-8711.2001.04006.x} multiplicity function, with the mass variance computed from the initial power spectrum alone, assuming Gaussianity and linear growth. The mass function at late times behaves as $M^{-0.6}$ for masses within the range $10^{-15}\lesssim M\lesssim 10^{-10}M_{\odot}$, which is comparable with N-body simulations.

By constructing merger trees we are able to estimate the redshift of collapse of a minicluster, and thus, following Navarro et al~\cite{Navarro:1996gj} estimate the halo concentration-mass relation, $C(M)$. The result is also relatively well described by using mass variance computed from the initial power spectrum assuming Gaussianity, and the analytic formulae of Navarro et al. A single calibration parameter, which is the same for the Peak Patch merger trees and the analytical model, gives $C(M)$ matching the N-body simulations for large masses.

Our method predicts that $C(M)$ reaches a maximum at $M\approx 10^{-13}(m_a/50\mu\text{eV})^{-0.5}M_\odot$, caused by the transition from direct collapse to hierarchical structure formation at matter radiation equality. This feature has not been observed yet in N-body simulations, which do not have the required small scale resolution.

For the box size of our initial conditions, the Peak Patch and N-body methods cannot be applied beyond $z=99$. However, since our analytical methods describe $C(M)$ well at large masses, we can use the linear power spectrum to extrapolate them to $z=0$, where we predict $C\sim \mathcal{O}(\text{few})\times10^4$ for typical miniclusters. This value of $C$ implies that typical miniclusters today have scale radii some order of magnitude larger than their Einstein radius for microlensing of stars in M31~\cite{Niikura:2017zjd}. Such typical miniclusters are expected to be much too diffuse to create microlensing signals ~\cite{Fairbairn:2017dmf,Fairbairn:2017sil}.

However, all is not lost for minicluster microlensing. Firstly, our present initial condition box is not large enough to describe the statistics of $C(M)$ and investigate the existence of the very rare and dense miniclusters that can have appreciable microlensing signals~\cite{Fairbairn:2017dmf,Fairbairn:2017sil}. If $C(M)$ has strong non-Gaussian tails, which is expected based on the distribution of initial overdensities~\cite{Kolb:1995bu,Vaquero:2018tib,Buschmann:2019icd} it may be possible for very rare and very dense miniclusters to form.

More promising is the possibility dense substructure whose density profiles are not well predicted by the NFW formalism adopted in this work. it is largely thought that NFW density profiles are formed primarily through hierarchical structure formation, that is, due to the mergers of many smaller halos. However, some of the sub-halos which formed very early (prior to matter-radiation equality) have had very few significant merger events and have instead grown primarily through the accretion of background matter, often referred to as self-similar in-fall. We expect that this will produce density profiles which are not NFW but instead perhaps more similar to the familiar 9/4 power-law. Such objects are able to produce microlensing signals. Therefore, combined with the large minicluster bound fraction, we believe these early forming sub-halos might be the key to observing axion miniclusters through microlensing.

While we have used the initial density field produced by simulations performed by Vaquero et al., there exist a number of other similar simulations. Each of these simulations differ slightly, incorporating different phenomena into the initial evolution of the axion field and/or simulating the fields evolution during different epochs. For example, the simulation method of Gorghetto et al. does not go through the $H=m_a$ epoch, and so cannot give the axion density field distribution itself . Instead, it is only suitable for estimating the relic density \cite{Gorghetto:2018myk}.

Other simulations such as those by Buschmann et al. and the slightly older results by Klaer \& Moore do go through the $H=m_a$ epoch and could therefore provide suitable density fields \cite{Buschmann:2019icd, Klaer:2017ond}. The published statistics of these simulations are in reasonable agreement with those of Vaquero et al. In the future, we plan to due the density fields produced by these simulations for comparison to the results presented here. While we do not expect the results to deviate significantly, any differences seen will be invaluable to our understanding of the role of the early evolutions of the axion field in the formation of axion miniclusters.

\acknowledgements{We are extremely grateful to Javier Redondo for providing the initial conditions from Ref.~\cite{Vaquero:2018tib}. We would also like to thank Benedikt Eggemeier and Javier Redondo for valuable discussions. DE and DJEM are supported by the Alexander von Humboldt Foundation and the German Federal Ministry of Education and Research. This research was supported by the Munich Institute for Astro- and Particle Physics (MIAPP) which is funded by the Deutsche Forschungsgemeinschaft (DFG, German Research Foundation) under Germany´s Excellence Strategy – EXC-2094 – 390783311. Lastly, we acknowledge the use of the open source Python packages NumPy \cite{5725236}, SciPy \cite{2019arXiv190710121V} and Matplotlib \cite{4160265}.}

\bibliographystyle{h-physrev3}
\newpage
\bibliography{axion}

\end{document}